\documentclass[12pt]{iopart}

\usepackage{cite}
\usepackage[pdftex]{graphics}
\usepackage{graphicx}
\DeclareGraphicsExtensions{.jpg,.pdf,.eps}
\usepackage{dcolumn}
\usepackage{bm}
\usepackage{iopams}
\usepackage{textcomp}	 
\usepackage{units}	 
\usepackage{mathptmx}

\renewcommand{\d}{\mathrm{d}}
\renewcommand{\i}{\mathrm{i}}

\renewcommand{\L}{\mathrm{L}}
\newcommand{\R}{\mathrm{R}}

\newcommand{\void}[1]{}

\begin{document}

\title[Theoretical and experimental investigations of Coulomb
blockade...]{Theoretical and experimental investigations of Coulomb blockade in
coupled quantum dot systems}

\author{F. J. Kaiser$^1$, S. Kohler$^1$, P. H{\"a}nggi$^1$, M. Malecha$^2$,
J. Ebbecke$^2$, A. Wixforth$^2$, H. W. Schumacher$^3$, B. K{\"a}stner$^3$ and 
D. Reuter$^4$ and A. D. Wieck$^4$} 
\address{$^1$ Theoretische Physik I, Institut
f{\"u}r Physik der Universit{\"a}t Augsburg, Universit{\"a}tsstr. 1, 86135
Augsburg, Germany} 
\address{$^2$ Experimentalphysik I,
Institut f{\"u}r Physik der Universit{\"a}t Augsburg, Universit{\"a}tsstr. 1,
86135 Augsburg, Germany} 
\address{$^3$ Physikalisch-Technische Bundesanstalt, Bundesallee 100, D-38116
Braunschweig, Germany} 
\address{$^4$ Angewandte Festk{\"o}rperphysik,
Ruhr-Universit{\"a}t Bochum, 44780 Bochum, Germany}
\ead{franz.josef.kaiser@physik.uni-augsburg.de}

\date{\today}

\begin{abstract}
Two strongly coupled quantum dots are theoretically and experimentally
investigated. In the conductance measurements of a GaAs based low-dimensional
system additional features to the Coulomb blockade have been detected at low
temperatures. These regions of finite conductivity are compared with theoretical
investigations of a strongly coupled quantum dot system and good agreement of
the theoretical and the experimental results has been found. 
\end{abstract}

\pacs{73.23.-b, 	
73.63,Nm, 		
73.63.Kv 		
}
\maketitle

\section{Introduction}

The electron transport through a single quantum dot in a two-terminal
configuration is governed by the interaction energy of the electrons
on the dot.  In many experiments, the state of the quantum dot is
essentially characterised by the electron number since orbital degrees
of freedom do not play a major role and can thus be ignored.  In the
limit of weak dot-lead coupling, the resulting current is determined
by states with an energy above the Fermi energy of one, but below the
Fermi energy of the other lead.  The other states suffer Coulomb
blockade: Energy conservation together with Pauli's exclusion
principle preserves their occupation number and, consequently, they
cannot contribute to the transport.  Only when the dot-lead coupling
becomes larger, co-tunneling processes start to play a role and
suspend Coulomb blockade.

When two or more quantum dots are in a linear transport arrangement between
two leads, the inter-dot tunneling can be incoherent or coherent,
depending on the coupling strength.  Incoherent tunneling is
sequential, i.e.\ between two tunneling events, the electrons dwell in
one particular dot.  Coherent tunneling is found for strong inter-dot
coupling such that the electrons reside in the delocalised
eigenstates of the double dot.  The analogy to $\pi$-electrons
in molecules is reflected by the term ``artificial molecule''.
A convenient theoretical picture for coherently coupled quantum
dots is a one single central system in which orbital degrees of
freedom play a role.

In an unbiased double dot, the relevant orbitals are the bonding and
the anti-bonding superposition of the localised states.  Then an
electron prepared in one dot will tunnel forth and back to the other
dot with a frequency set by the tunnel splitting.  These coherent
oscillations can be observed by lowering after a waiting time the
chemical potential of, say, the right lead.  If at that stage the
electron is in the right dot, it will tunnel to the right lead.
Periodic repetition of this procedure yields a dc current that
reflects the coherent oscillations \cite{Hayashi2003a}.  The coherence
of the superposition together with the possibility to perform a
readout allows one to devise charge qubits with double quantum dots.
The orbital degrees also influence the transport under microwave
excitation: Microwave irradiation can induce electron transitions from the
ground state to an excited state and thereby enhance the electron
transport between the leads, so that one observes photon-assisted
tunneling \cite{vanderWiel2003a, Platero2004a, Kohler2005a}.

A further common method to characterise low-dimensional semiconductor
systems such as the mentioned quantum dots, are conductance
measurements at low temperature: Since the current changes whenever an
energy level enters or leaves the voltage window, the differential
conductance exhibits a corresponding peak.  Shifting, in addition, the
energy levels by a gate voltage yields the characteristic ``Coulomb
diamonds'' which are observed in the differential conductance as a
function of gate voltage and bias voltage.
Within this work, we study both theoretically and experimentally the
fingerprints of orbital degrees of freedom in the Coulomb diamond
structure of coherently coupled quantum dots.  In
section~\ref{sec:experiment}, we describe our experimental setup and
present transport measurements, while in section~\ref{sec:theory}, we
study a minimal model that exhibits the observed Coulomb diamonds.
Moreover, we relate our theoretical findings to the experimental data.

\section{Experimental setup}
\label{sec:experiment}

\begin{figure} 
 \begin{center} 
 \includegraphics[width=8.0cm]{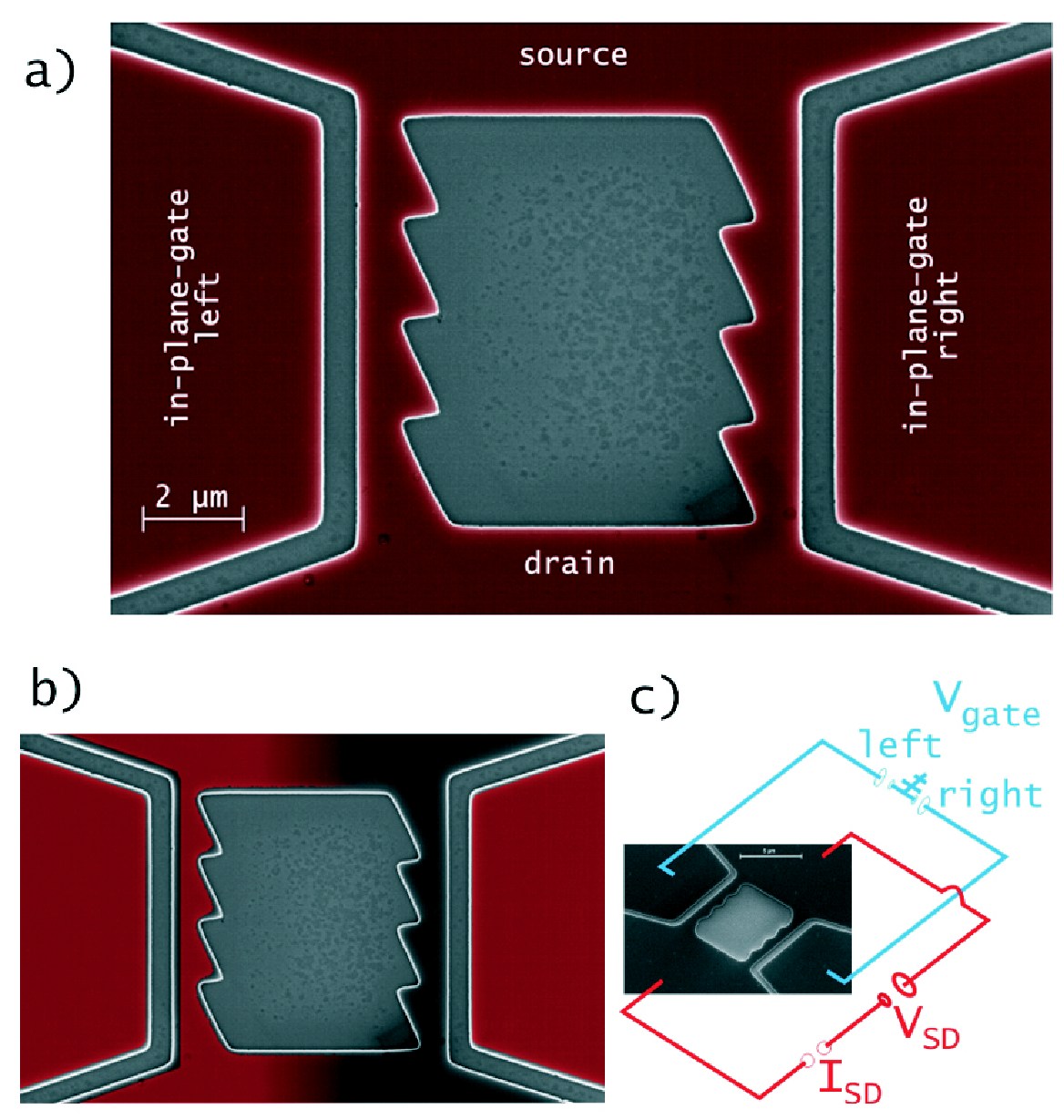}
 \end{center} 
 \caption{(a) Scanning electron micrograph of the sample. Red (darker)
  areas are highlighting 2DEG, black (brighter) areas are wet-etched and
  non-conducting. The annealed ohmic contacts (not shown) are labelled.
  The depletion of the left channel (under investigation) is controlled
  by the left in-plane gate. The channel can be made narrower by applying a
  more negative voltage to the gate until all electrons are forced out of the
  channel.
  In our experiment the right channel is completely depleted by a applying
  a sufficiently large voltage ($-4.5\,\mathrm{V}$) to the right gate
  as shown in panel (b).
 (c) Wiring scheme of the sample. }
 \label{fig:fig1} 
\end{figure}
\begin{figure}
 \begin{center}
  \includegraphics[width=8.0cm]{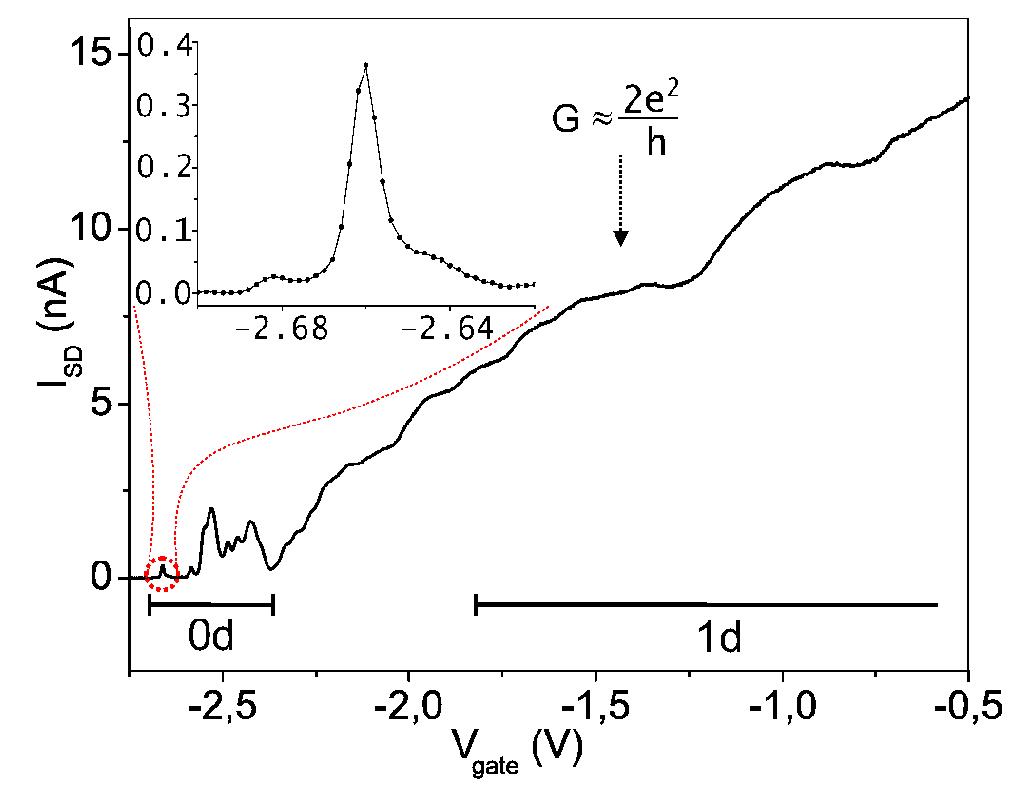}
 \end{center} 
 \caption{Measured current through the left channel with decreasing voltage on
 the left in-plane gate at $1.3\,\mathrm{K}$. The applied source-drain-voltage
 is $V_{\rm {SD}} =0.1\,\mathrm{mV}$. The system is evolving from the 1d regime,
 characterised by  current plateaus, the into QD-regime with characteristic
 Coulomb  oscillations. Inset: blow-up of the very first oscillation
 ($U_\mathrm{gate}=-2.66\,\mathrm{V}$) next to the pinch-off.} 
 \label{fig:fig2}
\end{figure}

%
In this work, we used a GaAs/AlGaAs heterostructure for sample
fabrication, where the two dimensional electron gas (2DEG) is located
approx.\ $55\,\mathrm{nm}$ beneath the sample
surface\cite{Davies2000a}. First, a mesa structure was defined by
photolithography, followed by wet-etching and annealing of ohmic
contacts for source, drain, and in-plane gates. Then the nanostructure
formation has been processed by electron beam lithography and
wet-etching\cite{Smith1996a}. Figure \ref{fig:fig1} shows the
nanostructure under investigation. While the grey (light) areas were etched,
the red (dark) areas depict the regions containing a high mobility 2DEG. The
structure is $8\,\mu\mathrm{m}$ long and in total about $10\,\mathrm{\mu m}$
wide. The inner structure is sawtooth shaped with four teeth and
asymmetric with respect to the vertical centre line. The conductive
channels between the sawtooth-tip and in-plane gates are approx.\
$0.9\,\mathrm{\mu m}$ wide. The 2DEG has an electron density of
$3.95\times10^{15}\mathrm{\mu m^{-2}}$ and a mobility of
$51.7\,\mathrm{m^{2}/Vs}$ (both measured in the dark at 
$T = 4.2\,\mathrm{K}$).  The sample allows us to carry out
measurements individually on each channel.  Here, we report only on
measurements on the left channel.  In order to ensure that only the
left channel is conductive and the measurement is not affected by the
right channel, a relatively high negative voltage of
$-4.5\,\mathrm{V}$ is applied to the right in-plane gate.
This causes a depletion\cite{Fowler1982a,Pepper1978a} of the 2DEG in
the right channel such that no electrons can pass from source to
drain. This situation is shown in figure~\ref{fig:fig1}(b), while
figure~\ref{fig:fig1}(c) sketches the wiring scheme of the sample.
In the same manner, the left channel can be depleted as well. If a
negative gate voltage is applied, the channel becomes narrower
until three potential barriers between the depleted areas, developed
from the voltage on the gate and the wet-etched tines, are formed.
Now, the 1d channel is separated into shorter channels and
with decreasing gate voltage it eventually evolves into small quantum
dots. At a certain voltage, the so-called pinch-off, the channel is
completely depleted and, consequently, no current can flow. Such an
evolution from 1d-channel \cite{Landauer1957a, Beenakker1991a} over
quantum dots (QD) \cite{Beenakker1991a, vanHouten1991a,
Livermore1996a, Ciorga2000a, Kouwenhoven1995a} to total depletion is
shown in figure~\ref{fig:fig2}. At less negative gate voltages,
characteristic 1d-conductance quantisation in the form of current
steps --- conduction plateaus --- can be seen. At even more negative
voltages, current oscillations were observed, being
characteristic for QD transport. The irregularity of the current peaks
as a function of the side gate voltage already indicates the existence
of a rather complicated electronic structure close to the pinch-off.
The measurement in figure \ref{fig:fig2} was carried out at
temperature $T=1.3\,\mathrm{K}$ with source-drain bias voltage of
$V_{\rm {SD}}0.1\,\mathrm{mV}$.

\begin{figure} 
 \begin{center}
  \includegraphics[width=8.0cm]{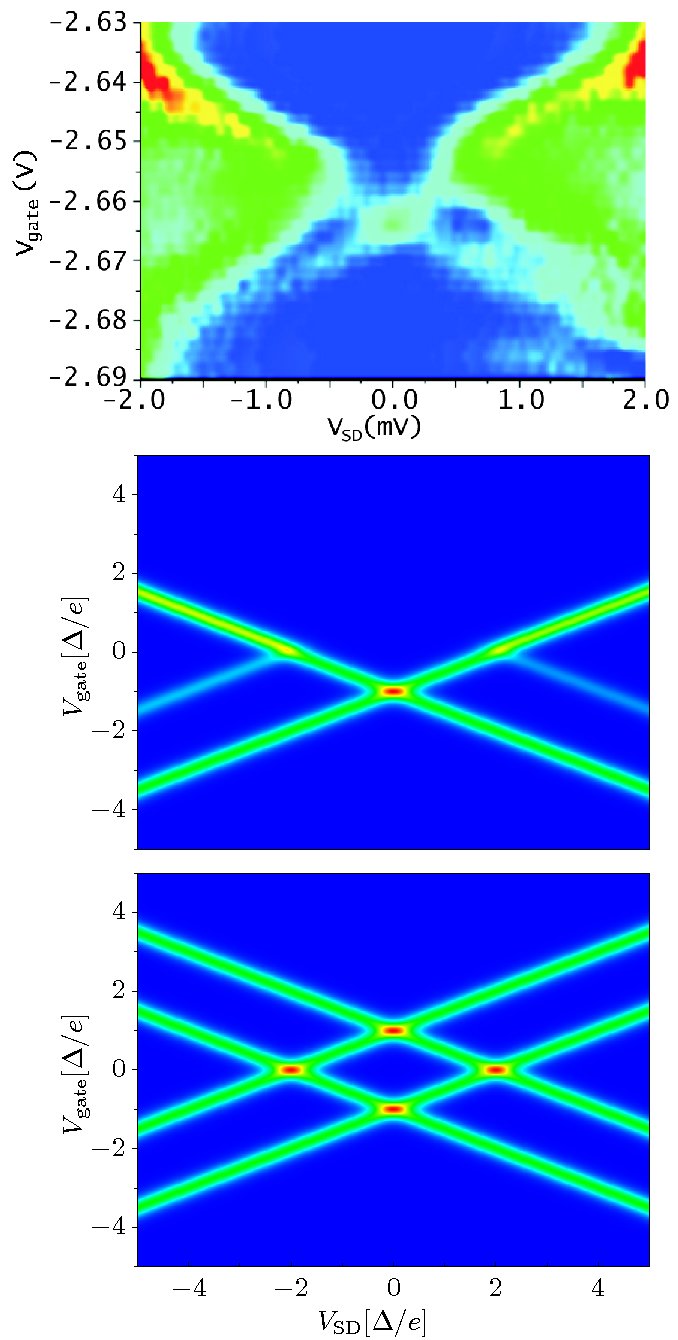}
 \end{center} 
 \caption{False-colour-plot of the differential conductance
  $\mathrm{d}I/\mathrm{d}V_\mathrm{SD}$ as a function of source-drain bias
  and gate voltage. Blue corresponds to low conductance and red to
  high conductance.  The experimental data (a) described in section
  \ref{sec:experiment} are compared to theoretical results for a double
  quantum dot with interacting (b) and non-interacting (c) electrons.
  The theoretical calculations are for dot-lead couplings
  $\Gamma_\L = 0.2\Delta$, $\Gamma_\R = 0.25\Delta$ and
  temperature $T=0.1\Delta$.
}
\label{fig:fig3} 
\end{figure}

\begin{figure}
 \begin{center} 
  \includegraphics[width=8.0cm]{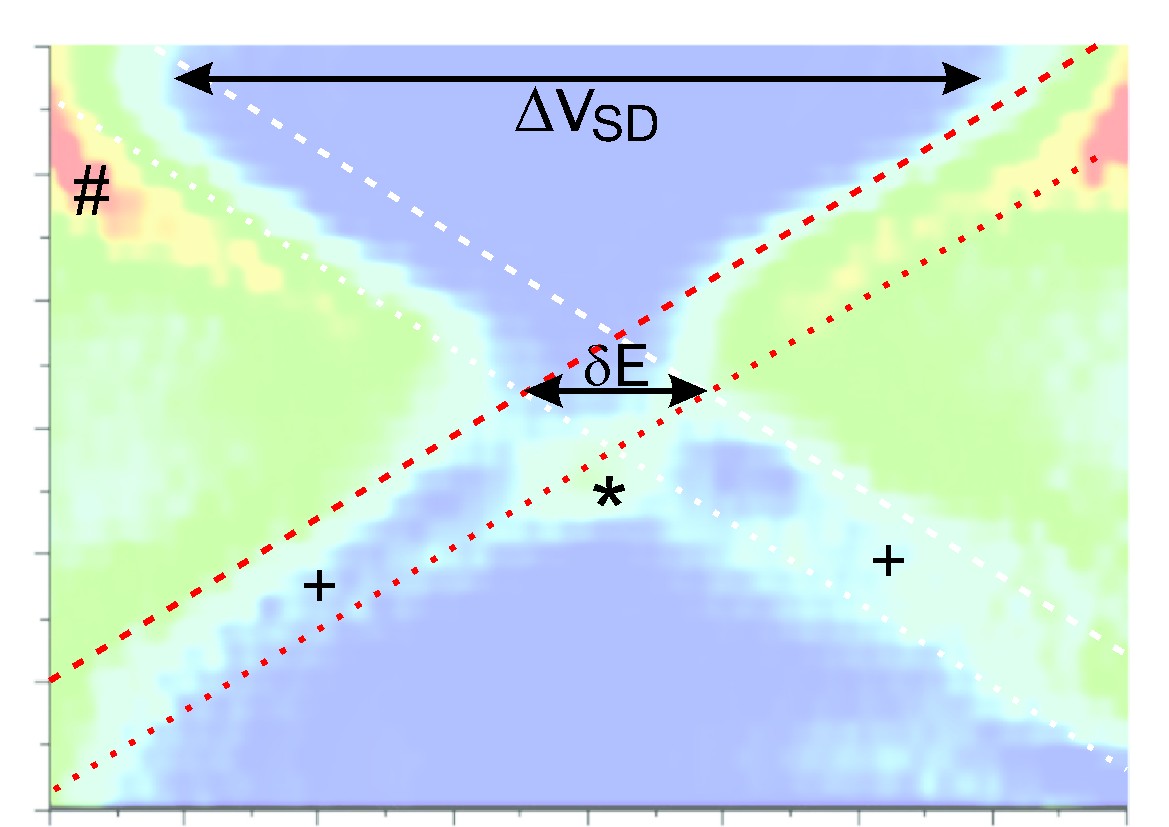} 
 \end{center} 
  \caption{Sketch of the system characteristics superimposed to the experimental
   data.From the slope of the dashed lines one can estimate the
   capacities of the dot, the doted lines highlight the stripes of
   elevated conductivity. The symbols *, + and \# mark parameter regions
   mentioned in the text. The total capacity of the system,
   $C_{\mathrm{\Sigma}}$, can be estimated from the value of $\Delta
   V_{\mathrm{SD}}$ \cite{Weis2003a}.}
 \label{fig:fig4} 
\end{figure}

Because it is quite difficult to estimate the exact landscape of the
barriers nor to explain the measurements in greater detail if the
channel consists of more than one quantum dot, we focused on
measurements very close to the pinch-off, where the first current
oscillation appears. The first current peak is at
$V_{\mathrm{gate}}=\unit[-2.66]{V}$; see inset of
figure~\ref{fig:fig2}. This quantum dot system can be characterised by
a set of current measurements for different source-drain voltages
\cite{Weis2003a}. Figure~\ref{fig:fig3} shows the corresponding
differential conductance $\mathrm{d}I/\mathrm{d}V_\mathrm{SD}$ which exhibiting
characteristic Coulomb-diamond structure \cite{Weis2003a,
vanHouten1991a, Ciorga2000a, Livermore1996a}.  From the slopes of the
transition from high to low conductivity shown in figure
\ref{fig:fig4}, one can extract the effective parameters which we use
later in our theoretical description \cite{Weis2003a}: the capacities
$C_{\mathrm{gate}}=(4\pm1)\times 10^{-18}\,\mathrm{F}$,
$C_{\mathrm{drain}}=(51\pm15)\times 10^{-18}\,\mathrm{F}$ and
$C_{\mathrm{\Sigma}}=(103\pm8)\times 10^{-18}\,\mathrm{F}$ where
$C_{\mathrm{gate}}$ is the capacity between the dot and the gate,
$C_{\mathrm{drain}}$ the capacity dot--drain, and
$C_{\mathrm{\Sigma}}$ the total dot capacity.  The single-electron
charging energy is $E_{\mathrm{C}}=0.78\pm 0.06\,\mathrm{meV}$ and the
energy spacing between two levels inside the dot is $\Delta E=1.6 \pm
0.1\,\mathrm{meV}$. The relatively large uncertainties 
follow from the estimated uncertainty of reading the quantities from
figure~\ref{fig:fig3} and from the calculated propagation of
uncertainty. If one assumes a parabolic potential, the energy spacing
lets us estimate the lateral dimension of $(14\pm
1)\times10^{-15}\,\mathrm{m^2}$ of the dot and the diameter of
$(136\pm 6 )\,\mathrm{nm}$. However, because of the triangular shape
from the lithographically defined sawtooth potential, it is unlikely
that the dot is perfectly round and a irregular shape is assumed.  

Figure \ref{fig:fig3}(a) shows the differential conductance
$\mathrm{d}I/\mathrm{d}V_\mathrm{SD}$.  It exhibits some differences
as compared to ``regular'' Coulomb diamonds. First, the diamond is
slightly tilted to the left. This is an indication for asymmetric
tunnelling barriers between the quantum dot and the leads as seen in
almost every Coulomb-diamond measurement. Second, some additional
structures in the diamond can be spotted.  The most obvious one is the
small area of high conductivity in the center section where the tips
of conductive areas almost merge (marked with * in
figure~\ref{fig:fig4}).  A further interesting feature is the narrow
stripes of finite conductance alongside the main areas (marked with
+). These mentioned areas are also rather symmetric due to bias
source-drain voltage and asymmetric along the gate voltage, i.e. they
appear only for lower gate voltage whereas the transition from
electron transport to Coulomb blockade for higher gate voltages is
very sharp. The third area (marked with \#) exhibits a high conductivity
and is also symmetric in the source-drain voltage but not as a
function of the gate voltage.  

\section{Theoretical descripton}
\label{sec:theory}

For a master equation description of electron transport at very low
bias voltages, one needs to take particular care in order to avoid
inconsistencies like the emergence of spurious non-vanishing transport
in equilibrium situations.  Such problems typically arise from the
approximation in the
interaction representation of the coupling operator
\cite{Novotny2002a, Kohler2005a}.  An detailed derivation of such a
master equation approach has been presented e.g.\ in reference
\cite{Kaiser2006a}.  Here, we will briefly review this approach.

\subsection{Model} 
\label{sec:model}

\begin{figure}[t]
  \centering
  \includegraphics[width=8.0cm]{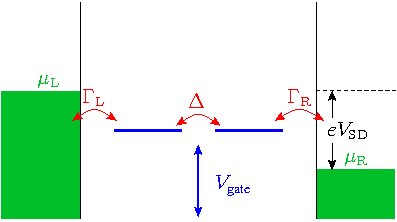}
  \caption{Tight-binding model for a double quantum dot coupled to two
  leads. An external bias voltage $V=(\mu_{\R}-\mu_{\L})/e$ is
  applied to the mesoscopic system.}
  \label{fig:model.channel}
\end{figure}%
The setup at hand for studying coherent quantum transport is shown in
figure~\ref{fig:model.channel} and the corresponding Hamiltonian reads
\begin{equation}
  \label{eq:H}
  H = H_{\mathrm{dots}} + H_{\mathrm{leads}} + H_{\mathrm{dots-lead}}.
\end{equation}
The individual terms describe the quantum dots, the electron
reservoirs of the leads, and the coupling of the dots to the leads. The
system itself is treated in a tight-binding approximation which we
restrict in the following to two orbital degrees of freedom. Since we
aim at exploring blocking effects, the corresponding wire Hamiltonian,
incorporating the Coulomb repulsion in the limit of a large
interaction strength $U$, assumes the form
\begin{equation} 
\label{eq:Hdots} 
H_{\mathrm{dots}} = \sum_{n} E_{n} c_n^\dagger
c_{n} - \Delta  \left(c_{2}^\dagger c_{1} + c_{1}^\dagger c_{2}\right) + U
\mathcal{N}(\mathcal{N}-1).  
\end{equation} 
The fermion operators $c_n^\dagger$ ($c_{n}$) create (annihilate) an
electron in the orbital $|n\rangle$ and $E_{n}$ denotes the respective
on-site energy.  In the Coulomb interaction term,
$\mathcal{N}=\sum_{n} c_n^\dagger c_{n}$ is the operator counting the
excess electrons on the dots. The inter-dot coupling is characterised
by the hopping matrix element $\Delta$. The leads attached to the dots
are modelled by ideal Fermi gases, 
\begin{equation} 
\label{eq:Hleads}
H_{\mathrm{leads}} = \sum_{\ell=\L,\R} \sum_q \epsilon_{q} c_{\ell q}^{\dag}
c_{\ell q} , 
\end{equation} 
where $c_{\ell q}^\dagger$ ($c_{\ell q}$) creates (annihilates) an electron with
energy $\epsilon_{q}$ in lead $\ell=\L,\R$. As an initial condition, we employ
the grand-canonical ensemble of the electrons in the leads at inverse
temperature $\beta=1/k_{\mathrm{B}}T$ and with electro-chemical potentials
$\mu_{\L/\R}$. Therefore, the lead electrons are described by the equilibrium
Fermi function $f_{\ell}(\epsilon_{q})=\{1 + \exp[-\beta(\epsilon_{q} -
\mu_{\ell})]\}^{-1}$. For the initial density matrix, we then have
\begin{equation}
\label{eq:rholeadeq}
\varrho_\mathrm{leads,eq}\propto
\exp\left[{-\beta(H_\mathrm{leads}-\mu_\L N_\L -\mu_\R N_\R)}\right],
\end{equation} 
where $N_{\ell}=\sum_{q} c^\dagger_{\ell q} c_{\ell q}$ denotes the
electron number in the left and right lead, respectively. From this
follows that all expectation values of the lead operators can be
traced back to the expression 
$
\langle c_{\ell' q'}^\dagger c_{\ell q}\rangle =
\delta_{\ell\ell'} \delta_{qq'} f_{\ell}(\epsilon_{q}).  
$
The two dots couple via the tunnelling matrix element $V_{\ell q}$ to the state
$|\ell q\rangle$ in the respective lead. The Hamiltonian describing this
interaction has the form 
\begin{equation} 
\label{eq:Hdot-lead} 
H_{\mathrm{dot-lead}} =
\sum_{q} (V_{\L q} c^{\dag}_{\L q} c_{1} + V_{\R q} c^{\dag}_{\R q} c_{2}) +
\mathrm{H.c.} 
\end{equation} 
It will turn out that the influence of the tunnelling matrix elements is
completely characterised by the spectral density
$
\Gamma_\ell(\epsilon) = 2\pi\sum_q
|V_{\ell q}|^2 \delta(\epsilon-\epsilon_q) 
$
which becomes a continuous function of $\epsilon$ if the lead modes
are dense.  If all relevant lead states are located in the centre of
the conduction band, the energy-dependence of the spectral densities
is not relevant and can be replaced by a constant,
$\Gamma_\mathrm{L/R}(\epsilon) = \Gamma_\mathrm{L/R}$.  This defines the
so-called wide-band limit.

\subsection{Master equation approach}
\label{sec:current}

The computation of stationary currents can be achieved by deriving a master
equation for the dynamics of the dot electrons.  Thereby, the central idea is to
consider the contact Hamiltonian \eref{eq:Hdot-lead} as a pertubation.  From the
Liouville-von Neumann equation $\i\hbar \dot \varrho=[H,\varrho]$ for the total
density operator $\varrho$ one obtains by standard techniques \cite{May2004a}
the approximate equation of motion
\begin{equation}
\label{eq:mastereq}
\fl
\dot\varrho(t)
=  -\frac{\i}{\hbar}[H_{\rm dots}(t)+H_\mathrm{leads},\varrho(t)] 
   -\frac{1}{\hbar^2}\int_0^\infty \d\tau [H_\mathrm{dot-lead},
     [\widetilde H_\mathrm{dot-lead}(-\tau),\varrho(t)]] .
\end{equation}
The tilde denotes operators in the interaction picture with respect to
the central system and the lead Hamiltonian, $\widetilde X(t) =
U_0^\dagger(t)\,X\,U_0(t)$, where $U_0$ is the propagator without the
coupling.  
The stationary current defined as the net (incoming minus outgoing)
electrical current through contact $\ell$ is given by minus the
time-derivative of the electron number in that lead multiplied by
the electron charge $-e$, $I_\mathrm{L}(t) = e(\d/\d t)\langle N_\ell\rangle$.
From the master equation \eref{eq:mastereq} follows
\begin{equation}
\label{eq:current.general}
I_\mathrm{L}(t)
= {}  e \tr[\dot \varrho(t) N_\mathrm{L}]
= {}
   -\frac{e}{\hbar^2} \int_0^\infty \d\tau \big\langle
   [\widetilde H_\mathrm{dot-lead}(-\tau),[H_\mathrm{dot-lead}, N_\mathrm{L}]]
   \big\rangle.
\end{equation}

In the following, we specify the master equation \eref{eq:mastereq}
and the current formula \eref{eq:current.general} for studying two
limiting cases: The first limit $U=0$ describes non-interacting
electrons.
The second limit refers to strong Coulomb repulsion such that $U$
is much larger than any other energy scale of the problem.  Then, only
the states with at most one excess electron on the wire are relevant.

\subsubsection{Non-interacting electrons}

In general, the relation between the states $|\phi_\alpha\rangle$ and
the many-particle Hamiltonian \eref{eq:H} is established via the
Slater determinant.  Alternatively, one can resort to Green's
functions.  In the present case, knowledge of the Green's function at
time $t=0$ is already sufficient. Apart from a prefactor, it is given
by the expectation value $P_{\alpha\beta}
=\langle c_\beta^\dagger c_\alpha\rangle$ for which
one obtains from equation~\eref{eq:current.general} for the stationary
current the expression
\begin{equation}
\label{eq:current0}
I_0 = \frac{e\Gamma_\ell}{\hbar} \sum_\alpha\Big[
\sum_\beta
 \langle \phi_\beta | n_\ell\rangle\langle n_\ell|\phi_\alpha \rangle
 P_{\alpha\beta} -
  |\langle n_\ell|\phi_\alpha\rangle|^2
  f_\ell(\epsilon_\alpha)
\Big] ,
\end{equation}
where the index $0$ refers to $U=0$.
It can be shown that the current is independent of the index $\ell$, i.e.\
independent of the contact at which it is evaluated.  This reflects for a
two-probe setting the validity of the continuity equation.
For the steady state expectation values $P_{\alpha\beta}$, we obtain from
the master equation \eref{eq:mastereq} the condition
\begin{eqnarray}
\label{eq:master0}
\fl
\i (\epsilon_\alpha -  \epsilon_\beta) P_{\alpha\beta}
=
\sum_{\ell=\L,\R} \frac{\Gamma_\ell}{2}
  \Big\{  
  \langle\phi_\alpha|n_\ell\rangle\langle n_\ell|\phi_\beta\rangle\,
  \big[ f_\ell(\epsilon_\alpha) + f_\ell(\epsilon_\beta) \big]
  \nonumber \\
  - \sum_{\alpha'}
  \langle\phi_\alpha| n_\ell\rangle\langle n_\ell|\phi_{\alpha'}\rangle\,
  P_{\alpha'\beta}
  - \sum_{\beta'}
  \langle\phi_{\beta'}| n_\ell\rangle\langle n_\ell|\phi_\beta\rangle\,
  P_{\alpha\beta'}
  \Big\} .
\end{eqnarray}
In a non-equilibrium situation, the solution of this set of equations
generally possesses non-vanishing off-diagonal elements, which in some
cases turn out to be crucial.

\subsubsection{Strong Coulomb repulsion}

In the limit of strong Coulomb repulsion, $U$ is assumed to be so
large that at most one excess electron resides on the system.  Thus,
the available Hilbert space is restricted to the states
$\{ |0\rangle, c_\alpha^\dagger|0\rangle \}_{\alpha=1, 2}$, such that the
density operator can be written as
\begin{equation}
\rho = |0\rangle\rho_{00}\langle 0|
      + \sum_\alpha \big( c_\alpha^\dagger|0\rangle\rho_{\alpha0}\langle 0|
                         +|0\rangle\rho_{0\alpha}\langle 0|c_\alpha \big)
      + \sum_{\alpha\beta} c_\alpha^\dagger
      |0\rangle\rho_{\alpha\beta}\langle 0| c_\beta.
\end{equation}
while the current expectation value \eref{eq:current.general} becomes
\begin{equation}
\label{eq:current.inf}
I_\infty=e  \Gamma_\ell\sum_{\alpha} \big[ \sum_{\beta}
        \langle \phi_{\beta}| n_\ell \rangle \langle n_\ell | \phi_{\alpha}\rangle
        \bar{f}_\ell (\epsilon_{\alpha})
        \rho_{\alpha\beta}- |\langle\phi_{\alpha}| n_\ell \rangle |^2
        f_\ell(\epsilon_{\alpha})\rho_{00} \big] ,
\end{equation}
where $\bar f=1-f$.
The decomposition of the master equation \eref{eq:mastereq} into the
single-particle states $c_\alpha^\dagger|0\rangle$ provides for the stationary
state the set of equations
\begin{eqnarray}
\label{masterinf}
\fl
\i (\epsilon_\alpha - \epsilon_\beta) \rho_{\alpha\beta} =
\sum_{\ell=\L,\R}\frac{\Gamma_\ell}{2}\Big\{
        \langle \phi_{\alpha}| n_\ell \rangle \langle n_\ell | \phi_{\beta}\rangle
        \big(f_\ell(\epsilon_\alpha)+f_\ell(\epsilon_\beta)\big)\rho_{00}
	\nonumber \\ 
- \sum_{\alpha'}
        \langle \phi_{\alpha}| n_\ell \rangle \langle n_\ell | \phi_{\alpha'}\rangle
        \bar{f}_\ell(\epsilon_{\alpha'})\rho_{\alpha'\beta}
- \sum_{\beta'}
        \langle \phi_{\beta'}| n_\ell \rangle \langle n_\ell | \phi_{\beta}\rangle
        \bar{f}_\ell(\epsilon_{\beta'})\rho_{\alpha\beta'} \Big\} .
\end{eqnarray}
In order to fully determine the density operator, we need in addition
an expression for $\rho_{00}$ which can also be derived from the
master equation.  A more convenient alternative is provided by the
normalisation condition $\tr\rho = \rho_{00}+\sum_\alpha
\rho_{\alpha\alpha} = 1$.  For the sake of completeness, we remark
that from the master equation~\eref{eq:mastereq} follows $\rho_{\alpha0} =
\rho_{0\alpha} = 0$ in the stationary state.

\subsection{Comparison with experimental data}

\begin{figure}[t]
  \centering
  \includegraphics[width=7.0cm]{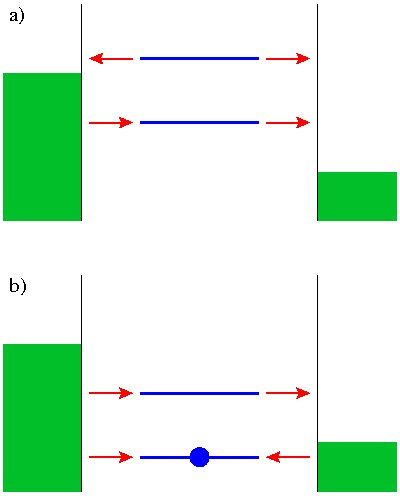}
  \caption{Sketch of the tranport through eigenenergy levels for
  different gate voltages. The arrows indicate the possible tunnel
  evenyts for electrons into and out of of the system. In panel (a),
  $V_{\rm Gate}$  is so large that only one level lies within the
  voltage window, while the other one lies well above and is never
  occupied.  Consequently, transport is interaction independent.  If one
  level lies below both chemical potentials~(b), it will occupied in the
  steady state and, thus, strong Coulomb repulsion inhibits the electron
  transport.}
  \label{fig:explain}
\end{figure}

Before establishing a quantitative relation between our model and the
experimental results, we discuss the transport properties of the
double-dot model qualitatively.  Thereby we reveal that both Coulomb
repulsion and an orbital degree of freedom play a role for the
behaviour for fixed, not too small source-drain voltage while the gate
voltage is changed.
For very large negative values of $V_{\rm gate}$, both eigenstates of
the double dot lie well above the chemical potential of both leads
and, thus outside the voltage window.
This means that lead states being in resonance with the dot states
remain unoccupied such that electron transport can only occur via
co-tunnelling processes. Thus, the current will be rather small.
When $V_{\rm gate}$ becomes larger such that the lower dot level lies
within the voltage window [see figure~\ref{fig:explain}(a)], resonant
transport becomes possible yielding a noticeable current.
Increasing the voltage further such that also the second level enters the
voltage window, opens a second path for non-interacting electrons 
through the dots. In the case of
strong Coulomb repulsion, however, double occupation of the dot is
impossible and, thus, the second orbital cannot fully contribute to
the transport.  Accordingly, the increase of the current is smaller.

The most significant difference between the two cases is found when
only the upper level lies within the voltage window, while the lower
level is below both chemical potentials, as sketched in
figure~\ref{fig:explain}(b).  Then the stationary state is
characterised by an occupied lower level.  Whether or not a further
electron can enter and cause a non-vanishing current now depends on
the strength of the Coulomb repulsion---for strong repulsion,
transport is Coulomb blocked. Consequently, for the two limits under
investigation, we obtain a current only in the one of non-interacting
electrons.  This is visible as an even qualitative difference in the
Coulomb diamond structure of figure~\ref{fig:fig3}:
The scenario for non-interacting electrons complies with particle-hole
symmetry.  This has the consequence that the corresponding Coulomb
diamond [figure~\ref{fig:fig3}(c)] is invariant under changing
the sign of both the source-drain voltage and the gate voltage.
For strong Coulomb repulsion, by contrast, the symmetry concerning the
sign of $V_\mathrm{gate}$ is no longer present; see
figure~\ref{fig:fig3}(b).  In particular for $V_\mathrm{SD} \approx
0$, the experimental data exhibit only one spot with high conductance,
which is in clear contrast to the theoretical result for
non-interacting case shown in figure~\ref{fig:fig3}(c).

The limits of strong and zero interaction have in common that when the
two energy levels enter the voltage window from above, the current
increases in two steps whose separation is determined by the energy
splitting $\delta E$ of the two levels, i.e.\ they are separated by
the voltage $\delta E/e$.  This enables one to determine the
excitation energy of an orbital degree of freedom by conductance
measurements.  Moreover, the source-drain voltage relates to the
steepness of the triangle.  Figure~\ref{fig:fig4} shows again the
measurement of figure~\ref{fig:fig3}(a) but now with the idealised
diamond structure marked by dashed lines.

For a more quantitative treatment, we compare the structure of the
experimental result in figure~\ref{fig:fig3}(a) with the theoretical
Coulomb diamond in figure~\ref{fig:fig3}(b).  This allows one to read
off the tunnel coupling $\Delta =\unit[0.8]{meV}$.  For the slightly
asymmetric lead-dot couplings $\Gamma_\L = 0.2 \Delta $ and $\Gamma_\R
=0.25\Delta $, we obtain at the plateaus for the current the values
\unit[2.2]{nA} and \unit[2.4]{nA}, respectively, which is of the order
of the measured values at the edges of the Coulomb diamonds ($\unit[2]{nA}-
\unit[3]{nA}$).

The quantitaive agreement between the experiment and the theoretical
result for $U=\infty$ suggests that electrons in the relevant
localised states of our sample strongly repell each other.  This rises
the question in which part of the sample (see figure~\ref{fig:fig1})
the localised states are formed.  By the chemical wet etching, the
sawtooth pattern has been created and four narrow constrictions
intersect the long wire into three separate regions. In
figure~\ref{fig:fig2}, features of a one-dimensional system have been
detected for less negative side gate voltages as discussed in an
earlier section. This means that one of the narrower constrictions
must govern the conduction process in the open channel regime, because
it is unlikely, that all constrictions represent identical tunnel
barriers.  However, we cannot identify which constriction dominates.
Moreover, in samples like the one used in this work, randomly
distributed charged impurities from the doping process are present.
They can strongly influence the potential profile depending on their
position during the cooling down process \cite{Nicholls1993a}.

Unfortunately, without further investigation, we are not able to
determine in which part of the sample the relevant levels are
localised.  With the data from the theoretical model for strong
inter-dot coupling, we nevertheless can infer, that both dots must be
rather close.  For the curved shape of the etched potential into
account, a unintentionally emerged dot is very likely as well.
 
\section{Conclusion}

We have studied Coulomb oscillations on lateral fabricated quantum
dots near the pinch-off.  In order to gain a theoretical
understanding, we investigated a two-site model which implies the
consideration of one orbital excitation.  A comparison of the measured
Coulomb diamond with theoretical predictions indicate that both
Coulomb repulsion and orbital degrees of freedom play a significant
role for the transport.  The importance of Coulomb repulsion is
emphasised by the fact that the corresponding model with
non-interacting electrons makes qualitatively wrong predictions.
Moreover, the theoretical results allow us to gauge the gate voltage
and to determine the energy splitting associated with the orbital
excitation.  As a drawback, the measurement does yield
any conclusion about the nature and the location of the two relevant
states. Studying a sample in which more orbital degrees of freedom
play a dominant role might provide additional information.  Such
experiments should be accompannied by theoretical studies for finite
Coulomb repulsion strength.

\section*{Acknowledgments}

Financial support of the German Excellence Initiative
via the ``Nanosystems Initiative Munich (NIM)'' and of the
Elite Network of Bavaria via the International
Doctorate Program ``NanoBioTechnology'' is gratefully acknowledged.
This work was supported in part by the Deutsche Forschungsgemeinschaft
under contract number EB 365/5.

\section*{References}
\bibliographystyle{jpcm.bst}

\end{document}